\documentstyle[aps,prl,multicol,epsf]{revtex}
\def\be{\begin{equation}}
\def\ee{\end{equation}}

\def\bea{\begin{eqnarray}}
\def\eea{\end{eqnarray}}
\begin{document}
\draft
\title{Width Distributions and the Upper Critical Dimension of KPZ Interfaces}

\author{E. Marinari${}^1$, A. Pagnani${}^2$, G. Parisi${}^1$,
and Z. R\'acz${}^{3}$}

\address{{}$^1${Dipartimento di Fisica, INFM and INFN, Universit\`a di Roma 
{\em La Sapienza}, P. A. Moro 2, 00185 Roma, Italy}}
\address{{}$^2${Dipartimento di Fisica and INFM, Universit\`a di Roma 
{\em La Sapienza}, P. A. Moro 2, 00185 Roma, Italy}}
\address{{}$^3$Institute for Theoretical Physics,
E\"otv\"os University, 1117 Budapest, P\'azm\'any s\'et\'any 1/a, Hungary}
\address{E-mail: Enzo.Marinari@roma1.infn.it, Andrea.Pagnani@roma1.infn.it,} 
\address{\hskip -28pt Giorgio.Parisi@roma1.infn.it, Racz@poe.elte.hu}
\date{May 9, 2001}

\maketitle

\begin{abstract}
Simulations of restricted solid-on-solid growth models are used 
to build the width-distributions of $d=2-5$ dimensional KPZ interfaces. 
We find that the universal scaling function associated with 
the steady-state width-distribution changes smoothly as $d$
is increased, thus strongly suggesting that $d=4$ is not an 
upper critical dimension for the KPZ equation. 
The dimensional trends observed in the
scaling functions indicate that the upper critical dimension 
is at infinity.
\end{abstract}

\date{\today}
\begin{multicols}{2}
\narrowtext

The KPZ equation \cite{KPZ} has been introduced to 
model growth in terms of a moving interface. The 
equation is written for the height $h({\vec r},t)$ 
of the interface above a $d$-dimensional substrate
\be
\partial_t h = \nu {\vec \nabla}^2 h +
\lambda ({\vec \nabla}h)^2 +\eta \, ,
\label{KPZeq}
\ee
where $\nu$ and $\lambda$ are  
parameters, while $\eta({\vec r},t)$ is a Gaussian white noise. 
Eq. (\ref{KPZeq}) can also give account of a number of other 
interesting phenomena (Burgers turbulence, directed 
polymers in random media, etc.) and, accordingly,  
a lot of efforts has been spent on finding and understanding 
the scaling properties of its solutions \cite{{Krugreview},{Barabasi},{HH1}}.
These intensive studies notwidthstanding, 
a number of unsolved issues remain, the question 
of upper critical dimension ($d_u$) being 
the most controversial one. 

The importance attached to $d_u$  
stems from the hope that, in analogy with equilibrium critical phenomena,  
a better understanding can be achieved through 
systematic expansions in terms of $d_u-d$. 
The search for $d_u$ has been on for about a decade 
\cite{{HH2},{Derr},{Schwa},{Cat},{Moo1},{Tu},{Moo2},{Lass},{Kert},{Kim},{MPP},{Piet}}
and the results range from $d_u\approx 2.8$ to $d_u=\infty$. 
Analytical estimates originate mainly from mode coupling theories  
which yield exact results for $d=1$ \cite{Frey}. 
Extending this approach to higher dimensions 
\cite{{Schwa},{Cat},{Moo1},{Tu},{Moo2}} one 
obtains values of $d_u$ which, after   
refining the self-consistency schemes,  
appear to settle to $d_u=4$. The result 
$d_u=4$ also emerges from various phenomenological field-theoretic 
schemes \cite{{HH2},{Lass}} and some nontrivial consequences of the
phenomenological arguments appear to be in agreement with 
simulations \cite{Nijs}.  

In contrast to the analytical approaches, numerical solution 
of the KPZ equation \cite{Kert}, simulations 
of systems belonging to the the KPZ universality class 
\cite{{Kim},{MPP}}, and the results of 
real-space renormalization group calculations 
\cite{Piet} provide no evidence for a finite $d_u$. 
Furthermore, the only numerical study \cite{Tu}
of the mode-coupling equations gives no indication for the 
existence of a finite $d_u$ either.

There are, of course, problems with both the analytic approaches 
and the numerical works. Assumptions about the 
scaling structure of the solution underlie the field
theoretic approaches, and uncontrolled approximations are
made when writing down the 
governing equation in mode-coupling theories. Additional 
uncertainties come from the use of various selfconsitency schemes 
in solving the mode-coupling equations. 
Simulations and numerical works have their own share of difficulties. The  
systems in higher dimensions cannot be large; the
extraction of exponents using fitting procedures which involve 
unknown correction-to-scaling exponents makes the error 
estimates suspect, and there may be difficulties with the 
numerical solution of the mode-coupling equations as well\cite{Moo2}.

In view of the above controversy, it is highly desirable to   
approach the $d_u$ problem in a way
unbiased by approximations and fitting 
procedures. Such an approach is described below where we study the 
steady-state width-distributions of $d=1$ to $d=5$ KPZ interfaces. 

The width-distributions have been introduced to provide a more detailed 
characterization of  
surface growth processes \cite{{FORWZ},{PRZ94},{RP94},{AR96}}, 
and they have been 
used to establish universality classes of rather divers phenomena 
\cite{{Korniss},{Tripathy},{BHF98},{Pinton1-2},{Holdsworth1-2},{turbu}}.
The quantity whose distribution is of interest here is  
the mean-square fluctuation of the interface defined by 
\be
w_2=\frac{1}{A_L}\sum_{\vec r} \left[\,h({\vec r},t)-\overline{h}
\,\right]^2\quad ,
\label{widthsq}
\ee
where $A_L$ is the area of the 
substrate of characteristic linear dimension $L$, and 
${\overline h}=\sum_{\vec r}h({\vec r},t)/A_L$
is the average height of the surface. Sampling $w_2$
in the steady state, one can build the so called 
width distribution $P_L(w_2)dw_2$,
defined as the probability that $w_2$ is in the interval $[w_2,w_2+dw_2]$. 
If the quantities $h(\vec{r},t)-\overline{h}$ were uncorrelated at
large distances the probability distribution of $w_2$ would be approaching
$\delta(w_2-\langle w_2\rangle_L)$ for $L\rightarrow\infty$. On the contrary, 
the fact that the
distribution is non trivial implies that these quantities are strongly
correlated at large distance.

The usefulness of this distribution lies in the following observation 
supported by all the examples studied so far 
(including $d=1$ and $2$-dimensional 
KPZ surfaces \cite{{FORWZ},{PRZ94},{RP94},{Holdsworth1-2}}).
Namely, in systems where the steady-state roughness diverges 
$\langle w_2\rangle_L\rightarrow \infty$ in the $L\rightarrow \infty$ limit, 
$P_L(w_2)$ assumes a scaling form  
\be 
 P_L(w_2)\approx \frac{1}{\langle w_2\rangle_{{}_L}} \Phi_d \left( 
 \frac{w_2}{\langle w_2\rangle_{{}_L}} \right) \quad ,
 \label{Phi}
\ee
where $\Phi_d (x)$ is an universal scaling function characteristic of the 
universality class of a given nonequilibrium dynamics in dimension $d$. This 
universality is understandable, it is a consequence of the facts that (i) 
a steady state can be considered as a critical state 
if the fluctuations diverge, and (ii) in critical systems, 
the distribution functions of macroscopic quantities 
(such as $\langle w_2\rangle$) are characterized 
by scaling functions which are universal.
  
The universality of $\Phi_d(x)$ allows 
the investigation of the problem of $d_u$, once it is noted  that 
the scaling functions depend on dimensionality up to $d=d_u$
and they are expected to take on a fixed shape for $d\ge d_u$. 
Thus if one finds that scaling functions vary smoothly 
in dimensions $1\le d \le {\hat d}$, one can conclude that ${\hat d}-1<d_u$.
This is the line of argument we employ below for KPZ systems.
We shall compare the scaling functions (\ref{Phi}) 
for $1\le d\le 5$ using the exact results for 
the $d=1$ steady state \cite{FORWZ}, previously obtained simulation data 
for $2\le d\le 4$ 
restricted solid-on-solid (RSOS) growth models \cite{MPP}, 
and by generating new data for the $d=5$ RSOS model. Our main 
finding is that the $\Phi_d(x)$-s change smoothly as $d$ is varied 
thus suggesting that $d_u>4$. 

It is important to recognize that
there are no fitting procedures in the above approach. 
The width-distributions are just histograms calculated from Monte-Carlo 
(MC) simulations. Both quantities $w_2$ and $\langle w_2\rangle_{{}_L}$ 
entering (\ref{Phi}) are measured and no scaling properties 
of  $\langle w_2\rangle_{{}_L}$ are used or assumed. 
The only approximation is the 
finite size of the systems investigated.
It should be noted, however, that our approach relies only on the 
shape of the scaling functions. 
Since the important size-dependences reside in the argument of these
functions the functional forms converge at small sizes. 
A further and rather important observation that helps to reach our 
conclusion is that, as we shall show below, the scaling functions converge 
to well distinguished forms for $1\le d\le 5$.

Let us now present and discuss the evidence for our conclusion of $d_u>4$. 
The scaling functions (\ref{Phi}) for dimensions $d=1-5$ are displayed 
in Figs. 1 and 2. The $d=1$ curve is an exact result \cite{FORWZ}.  
The rest is obtained from simulations of the RSOS model 
that is believed to belong to the universality class of the KPZ equation
\cite{unisos}. 
The RSOS model and its simulations are described in \cite{MPP} where
hypercubes of volume $L^d$ with
periodic boundary conditions were simulated and a multi-surface coding 
technique allowed to obtain excellent steady-state statistics 
for systems up $d=4$. We take the results from this work
to build $\Phi_d(x)$ for $d=2-4$ and, in this paper, 
we extend these simulations to find $\Phi_d(x)$  for $d=5$ as well.

\begin{figure}[htb]
\centerline{
        \vspace{-0.8cm}
        \epsfxsize=0.48\textwidth
        \epsfbox{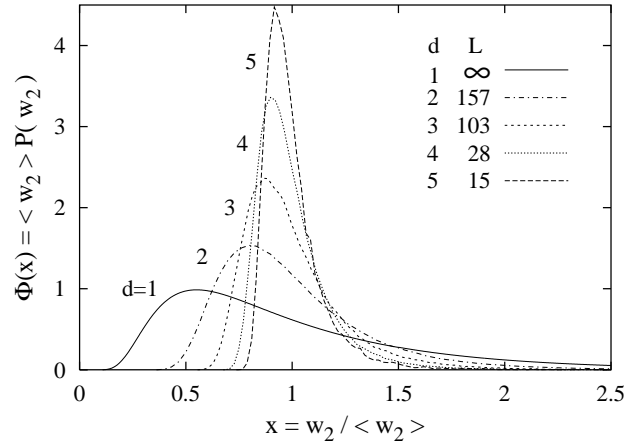}
           }
\vspace{1.2cm}
\caption{Scaling function for the width-distribution [eq. (\ref{Phi})] 
of KPZ systems in dimensions $d=1-5$, with $L$ giving the largest 
linear size of the system in which $\Phi_d(x)$ was measured.}
\label{Fig:d1-5}
\end{figure}

As one can see on Figs. 1 and 2, the scaling functions 
change smoothly as $d$ increases. The $\Phi_d(x)$-s get narrower and 
more centered on $x=1$, and there does not seem to be any break 
in this behavior at $d=4$. The 
equality of the d=4 and 5 scaling functions appears to be excluded.
Since our conclusion about $d_u>4$ rests on the above 
observations we must now discuss some details in order to make it
more than a visual observation. 

\begin{figure}[htb]
\centerline{
        \vspace{-0.8cm}
        \epsfxsize=0.48\textwidth
        \epsfbox{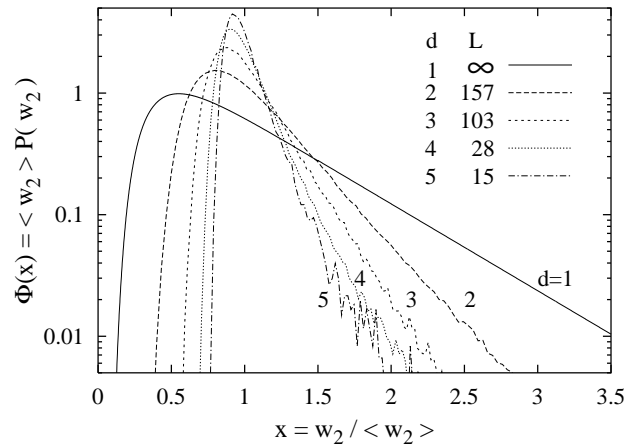}}
\vspace{1.2cm}
\caption{The same as Fig.1 but on a semilog scale in order to demonstrate the 
differences at small probabilities. Note that the range of $x$ has been enlargened 
so that the large-$x$ asymptotics would be better seen.}
\label{Fig:d1-5log}
\end{figure}
 
The basic problems that may arise in measuring steady state 
properties are the 
problems of statistics, relaxation, and finite-size. 
Since the multi-surface coding allowed the simulations 
of 32 or 64 systems in one run, we had no problem gathering data 
with good statistics. The relaxation time problems were taken care 
by having very long runs and being in the asymptotic plateau 
region of  $w_2$ for at least an order of magnitude longer period than 
the time of reaching the plateau (for details see discussion and Figs. 3, 5, 
and 7 in \cite{MPP}).  

\begin{figure}[htb]
\centerline{
        \vspace{-0.8cm}
        \epsfxsize=0.48\textwidth
        \epsfbox{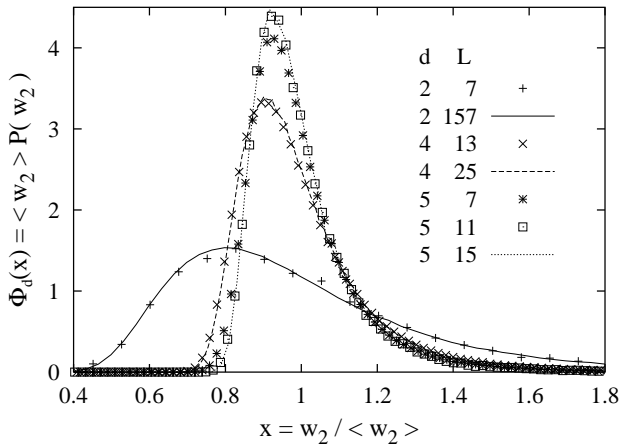}}
\vspace{1.2cm}
\caption{Finite-size effects on the scaling function [eq.(\ref{Phi})] in
dimensions $d=2,4$ and $5$. L denotes the linear size of the hypercubes
investigated.}
\label{Fig:finsize}
\end{figure}

The solution to the problem of finite size is less obvious.
In general, one can observe that the $\Phi_d$-s converge to their 
limiting shape 
when the number of sites ($N=L^d$) becomes about a factor 2 
larger than the number of sites on the surface ($N_s$). 
Fig.\ref{Fig:finsize} demonstrates this observation for 
the $d=2$ and $4$ systems. Results for linear sizes of $L=7$ 
($N\approx 2N_s$) and
$L=157$ ($N\approx 40N_s$) are compared for $d=2$ and one can see that they 
have the same $\Phi$-s within the
the statistical errors of the simulations \cite{staterr}. Similar conclusion 
can be drawn by comparing the $L=13$ ($N\approx 2N_s$) and 
$L=25$ ($N\approx 3.5N_s$) systems in $d=4$
(analogous results for $d=3$ are not displayed on Fig.\ref{Fig:finsize} 
in order to keep clarity in presentation).   

We need the $N\approx 2N_s$ convergence rule because
the largest $d=5$ system we can 
investigate has $L=15$, corresponding to 
$N\approx 2N_s$. The results for $\Phi_5$ displayed in 
Fig.\ref{Fig:finsize} indicate that the $N\approx 2N_s$
rule applies to $d=5$ as well. Indeed, systematic deviations 
between the $L=11$ and $L=15$ curves can be detected only at 
small values of $\Phi_5$ in the region of $x\le 0.85$.
An important feature of the size-dependence of $\Phi_5$ that can be seen 
in Fig.\ref{Fig:finsize} is that the 
maximum of $\Phi_5$ slightly increases with size. This means that, near the
maximum, $\Phi_5-\Phi_4$ becomes larger with increasing $L$, thus 
excluding the possibility of the two functions $\Phi_5$ and $\Phi_4$ 
becoming equal. 
The different functional forms for the scaling functions 
in $d=4$ and $5$ then indicate that $d=4$ is not the upper critical
dimension for the KPZ systems.

\begin{figure}[htb]
\centerline{
        \vspace{-0.8cm}
        \epsfxsize=0.48\textwidth
        \epsfbox{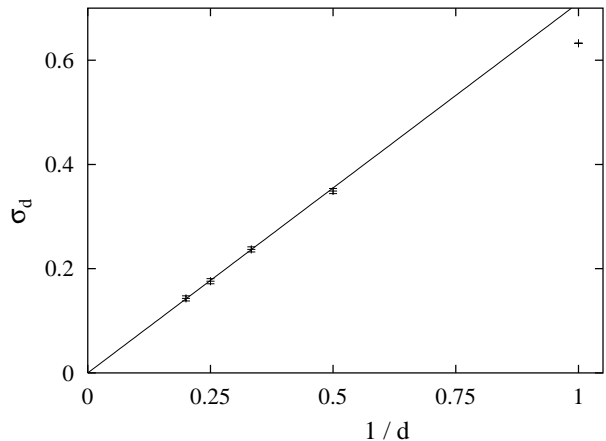}}
\vspace{1.2cm}
\caption{Dimensional dependence of the relative fluctuations of $w_2$. The
extrapolated values $\sigma_d(L\rightarrow\infty)=\sigma_d(\infty)$ 
[see eq.(\ref{sigma})] are plotted against $1/d$ with 
the solid line given by $\sigma_d= 0.71/d$.
}
\label{Fig:1overd}
\end{figure}

The concept of smooth changes across $d=4$ can be put on a more 
quantitative basis by examining the dimensional trend in the spread 
of the scaling function around its average $x=1$ 
\be
\sigma^2_d=\int_0^\infty dx \, (x-1)^2 \, \Phi_d(x) = 
\frac{\langle (w_2)^2\rangle}{\langle w_2\rangle^2} - 1 \, ,
\label{sigma}
\ee
that is related to relative mean-square fluctuations of $w_2$. 
Apart from the case of $d=1$, the $L$ dependence of $\sigma_d$ is
very weak and plotting $\sigma_d(L)$ against $1/L$ one can get accurate 
estimates of $\sigma_d(\infty)$. The values of $\sigma_d(\infty)$  
as a function of $1/d$ are displayed on Fig.\ref{Fig:1overd}.
As one can see, apart from the $d=1$ result,
the straight line $\sigma_d\approx 0.71/d$ gives an excellent 
description of the dimensional dependence of $\sigma_d$.

The result $\sigma_d\approx 0.71/d$ indicates that $\sigma_d\rightarrow 0$ 
for $d\rightarrow \infty$ i.e. the scaling function converges to a 
delta-function, $\delta(x)$, at $d=\infty$. 
Remarkably, the convergence $\Phi_d(x)\rightarrow \delta(x)$ 
also takes place in a related surface-growth model, 
in the Edwards-Wilkinson model if $d\rightarrow 2$ 
\cite{PRZ94}. Since $d=2$ happens to be the upper critical dimension 
of this model (the interface becomes flat for $d> d_u=2$) one may 
speculate that the results displayed on Fig.\ref{Fig:1overd} actually 
give support to the suggestion that failure of 
numerical attempts at locating a
finite $d_u$ means that $d_u=\infty$ for KPZ systems \cite{kpz5exp}.

In summary, we believe that the main results of this paper 
(Fig.\ref{Fig:d1-5} and \ref{Fig:d1-5log}) provides strong evidence that 
$d_u\ge 5$ and, furthermore, Fig.\ref{Fig:1overd} suggests that  
$d_u=\infty$. As a final remark, let us note that the results displayed on 
Fig.\ref{Fig:d1-5} and \ref{Fig:d1-5log} 
can be appreciated from another point of view.
Namely, we have constructed the scaling functions of width distributions
for the KPZ universality class. 
Thus we have expanded the picture gallery of 
scaling functions that may be used for identifying the universality 
classes of nonequilibrium steady states.


We thank G. Gy\"orgyi for useful discussions. 
This work has been supported by the 
Hungarian Academy of Sciences (Grant No. OTKA T 029792).


\end{multicols}


\begin{references}
\bibitem{unisos} The mapping onto the KPZ equation is supported not only by 
                 the comparison of critical exponents but also by the results
                 for the universal amplitude ratios [see e.g. J. Krug, 
                 P. Meakin, and T. Halpin-Healy, 
                 Phys. Rev. A{\bf 45}, 638 (1992)].
\bibitem{staterr} A non-negligible contribution to the statistical error 
                  comes from the fact that the number of possible $w_2$ 
		  values is small for small systems. This causes problems 
		  with the choice of binsize, $dw_2$, when building $P(w_2)$.
		  Namely, $dw_2$ must be significantly larger than the 
		  difference between neighboring $w_2$ values, otherwise 
		  small changes in $dw_2$ yield large fluctuations in $P(w_2)$.
		  On the other hand, $dw_2$ must be smaller than the scale 
		  on which relevant changes occur in $P(w_2)$. Finding 
		  appropriate values of $dw_2$ does not pose a problem 
		  for the system sizes we studied, in general. The $d=2$, 
		  $L=7$ system, however, is an exception where the 
		  fluctuations coming from binsize choice are clearly
		  visible.
\bibitem{kpz5exp} In addition to obtaining $\sigma_5$, the simulations of the 
                  $d=5$ KPZ system 
                  can also be used to determine the critical exponents in the
		  finite-size scaling of the width,
		  $w_2\approx A_2 L^{2\chi}(1+B_2L^{-\omega})$. One 
		  finds that $\chi = 0.205  \pm 0.015$ and 
		  $\omega =  1.07 \pm 0.14$. Although these results are 
		  obtained by fitting and thus are not used in the
		  main line of argument of the paper, it should be noted 
		  that, taken together with the results of the $d=2-4$ 
		  simulations \cite{MPP}, they give additional 
		  support to the notion of 
		  smoothly changing criticality across $d=4$. We 
		  also note that the $d=5$ simulations did not show any 
		  evidence of a glassy phase that has been discussed as
		  a possibility fo $d>4$ KPZ systems \cite{{Mezard},{Moo1}}.
\end{references}
\end{document}